\documentclass[a4paper,12pt]{article}
\usepackage[utf8]{inputenc}
\usepackage{cancel}
\usepackage{ulem}
\usepackage{amsfonts}
\usepackage{amssymb}
\usepackage{graphicx}
\usepackage{amsmath}
\usepackage{enumerate}
\usepackage{mathtools}
\usepackage{subfig}
\usepackage{color}
\usepackage{tikz}\usetikzlibrary{calc}
\usepackage{float}
\usepackage{here}
\usepackage{cite}
\usepackage{mathrsfs}
\usepackage{float,epsfig}
\usepackage{dcolumn}
\usepackage[toc,page]{appendix}
\usepackage{booktabs}   
\usepackage{multirow}   
\usepackage{graphicx}
\usepackage{bm}
\usepackage{amsmath,amssymb,amsthm}
\usepackage[colorlinks=true,linkcolor=blue,citecolor=red]{hyperref}
\newcommand{\be}{\begin{equation}}
\newcommand{\ee}{\end{equation}}
\newcommand{\bea}{\setlength\arraycolsep{2pt} \begin{eqnarray}}
\newcommand{\eea}{\end{eqnarray}}

\def\0{{\sst{(0)}}}
\def\1{{\sst{(1)}}}
\def\2{{\sst{(2)}}}
\def\3{{\sst{(3)}}}
\def\4{{\sst{(4)}}}
\def\5{{\sst{(5)}}}
\def\6{{\sst{(6)}}}
\def\7{{\sst{(7)}}}
\def\8{{\sst{(8)}}}
\def\sst#1{{\scriptscriptstyle #1}}

\makeatletter \@addtoreset{equation}{section}

\usepackage{multirow}
\definecolor{lime}{HTML}{A6CE39}

\tikzset{>=latex}

\begin{document}
\title{{\normalsize \textbf{\Large  Observational Aspects of Black Holes  in Square-Root Electrodynamic Backgrounds
 }
}}
\author{ {\small   Maryem   Jemri\thanks{maryem.jemri@um5r.ac.ma}
 \hspace*{-8pt}} \\
{\small  ESMaR, Faculty of Science, Mohammed V University in Rabat, Rabat, Morocco  }}
\maketitle

\begin{abstract}
Motivated by recent thermodynamic investigations \cite{nn}, we study the optical behaviors of black holes in a square-root nonlinear electrodynamic framework. More precisely, we examine the shadow aspect of black holes obtained by considering the electrodynamic Lagrangian  $\mathcal{L}(\mathcal{F})=
\mathcal{F} \sqrt{1-k(2Q^2\mathcal{F})^{1/4}},$ where $\mathcal{F}$ is the Maxwell gauge invariant,  $k$ is a relevant dimensionless  parameter, and $Q$ is considered as a magnetic charge. By confronting the obtained results with the Event Horizon  Telescope collaboration findings, we provide the corroborated moduli space regions. Then, we approach  such optical behaviors  using machine learning techniques. As a result, we observe a possible  compatibility with  the Sgr $A^*_{Keck}$  black hole  model. 
 {\noindent} 


\end{abstract}

%

\newpage

\section{Introduction}
Black holes have attracted a  considerable attention in connection with many  branches in  physics, particularly  modified theories of gravity \cite{i01,i02,3}. In this regard, nonlinear electrodynamics (NED) has been introduced as an interesting framework for examining the interplay between gravitational and electromagnetic fields beyond the standard Maxwell theory. NED is characterized by its field equations, which lead to modifications of the spacetime geometry and the associated black hole solutions \cite{i003,i03,i04,14,15,16,17,18,19}. Such corrections can affect various physical properties, including their thermodynamic and optical behaviors.\\
Beyond their thermodynamic properties, NED black holes have also been investigated from an optical aspect, particularly via their photon spheres, shadows, and gravitational lensing phenomena. The NED modifications can alter photon trajectories and  affect the geometric parameters  of black hole shadows  including  the  size and  the shape \cite{I1,I2,I3,I4}. Recently, the thermodynamic properties of black holes in modified theories of gravity have received considerable attentions \cite{i05,i06,i07,i08,i09}. Specifically, the extended thermodynamic approach has enabled the investigation of critical phenomena, phase transitions, and Van der Waals behaviors in black hole systems \cite{I5,I6,I7,I8,II9}. More recently, a new black hole solution in square-root nonlinear electrodynamics (SRNED) has been proposed, revealing Van der Waals  behaviors and $P$--$v$ criticality without  corrections  recovering directly  the  relevant universal ratios \cite{nn}. These  findings  have further motivated the investigation of the optical properties of black holes within the SRNED framework.\\
The aim of this work is to investigate the optical properties of non-rotating and rotating  black holes in SRNED backgrounds, focusing on their shadow curve configurations.  Precisely, we examine the effects of the  involved model parameters and confront the resulting observables with the EHT constraints for M87$^*$ and Sgr A$^*$ black hole models. Furthermore, we employ machine learning techniques  to identify the regions of the moduli space compatible with the observational data and assess the classification performance. The obtained  results reveal compatibilities  with the Sgr A$^*_{Keck}$ observations.\\
The organization of this paper is as follows. In section  2, we present a short discussion of black holes in SRNED backgrounds. Section  3 is devoted to the shadow computations of rotating and non-rotating SRNED black holes. In section 4, we assess the current theoretical predictions by comparing them with EHT observations. In section  5, we employ machine learning methods to classify whether the reduced parameter space of such SRNED black holes is consistent with the observational constraints of M87$^*$ and Sgr~A$^*$ models. Finally, section  6 provides the concluding remarks. 
\section{Rotating black holes  in SRNED backgrounds}
\begin{figure}[h!]
\begin{center}
	\includegraphics[scale=0.34]{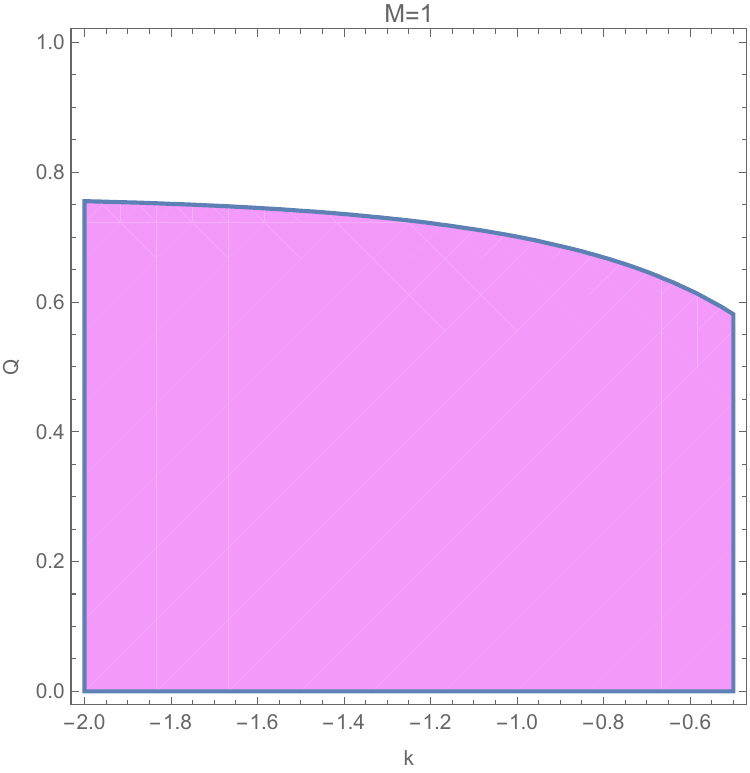}
\end{center}

\caption{ \it \footnotesize Parameter region of the moduli space ensuring the existence of a real event horizon for  the non-rotating solution.}
\label{H1}
\end{figure}

In this section, we reconsider the study of charged black holes in the context  of SRNED reported in \cite{nn} in order to examine   the optical behaviors in the next sections. To this end, we start from the action describing standard Einstein gravity  coupled to NED written as
\begin{equation}
S=\frac{1}{16\pi}\int d^4x\,\sqrt{-g}
\left[R+4\mathcal{L}(\mathcal{F})\right],
\end{equation}
where $R$ represents the Ricci scalar, and $\mathcal{L}(\mathcal{F})$ is an arbitrary function of the electromagnetic field invariant $\mathcal{F}=\frac{1}{4}F_{\mu\nu}F^{\mu\nu}$. Here, $F_{\mu\nu}=\partial_\mu A_\nu-\partial_\nu A_\mu$ denotes the electromagnetic field strength tensor.
To investigate four-dimensional black hole solutions, we assume a static and spherically symmetric spacetime defined by the following line element
\begin{equation}
ds^2=-f(r)dt^2+\frac{1}{f(r)}dr^2
+r^2d\theta^2+r^2\sin^2\theta\,d\phi^2,
\end{equation}
where $f(r)$ represents the metric function. As proposed in \cite{nn}, we adopt the following NED Lagrangian
\begin{equation}
\mathcal{L}(\mathcal{F})=
\mathcal{F} \sqrt{1-k(2Q^2\mathcal{F})^{1/4}},
\end{equation}
where $Q$ is a constant which could be interpreted as a magnetic charge, while  $k$ denotes a 
dimensionless parameter characterizing the SRNED model. For the electric field configurations, the electromagnetic invariant takes the form
\begin{equation}
\mathcal{F}=\frac{Q^2}{2r^4}.
\end{equation}
Solving the Einstein field equations within the SRNED formalism by neglecting the cosmological constant for the purpose of investigating the black hole shadow yields the following metric function
\begin{equation}
f(r)=1-\frac{2M}{r}
-\frac{2Q}
{3k r}\left(1-\frac{k Q}{r}\right)^{3/2}
\label{o}
\end{equation}
where $M$ denotes the black hole mass and $Q$ is the associated magnetic charge. To explore the physical properties of the reconsidered black hole solutions, we first investigate the existence of event horizons, which are determined by the  vanishing of the inverse radial component of the metric tensor. To examine the conditions under which the horizons exist, we fix the black hole mass to $M=1$ and investigate the effects of the parameters $k$ and $Q$ describing a two dimensional moduli space. Figure~(\ref{H1}) illustrates the physically admissible regions in such a  reduced moduli space.
As shown in this figure, negative values of $k$ significantly restrict the range of the charge $Q$ that allows the existence of an event horizon. Furthermore, as  $k$ increases, the regions in the $(Q,k)$ parameter space admitting real values of the horizon radius $r_h$  become smaller. This behavior further  constrains  the charge, which must satisfy $0 \leq Q \leq 0.75$.

\begin{figure}[h!]
\begin{center}
	\includegraphics[scale=0.3]{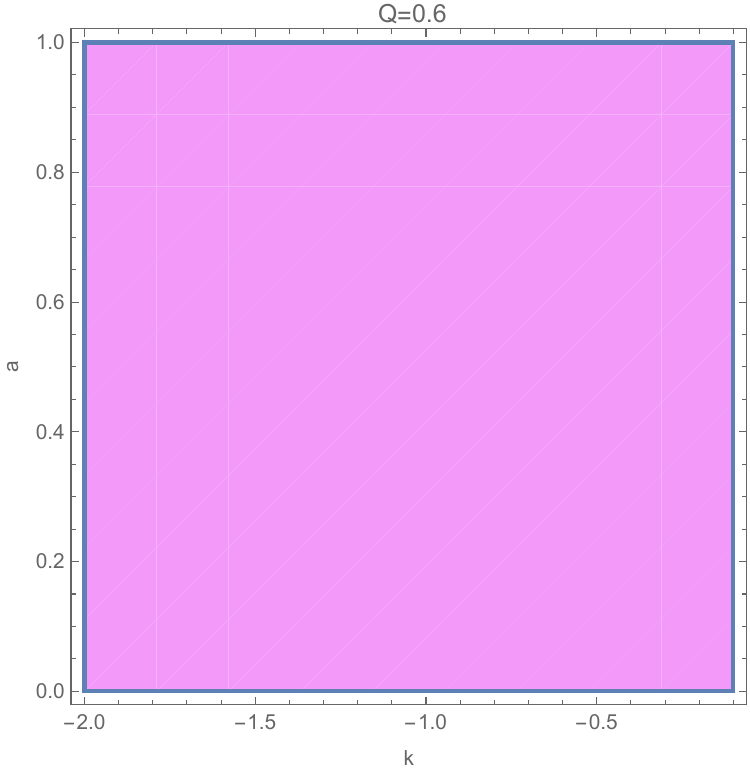}\hspace{1cm}
	\includegraphics[scale=0.3]{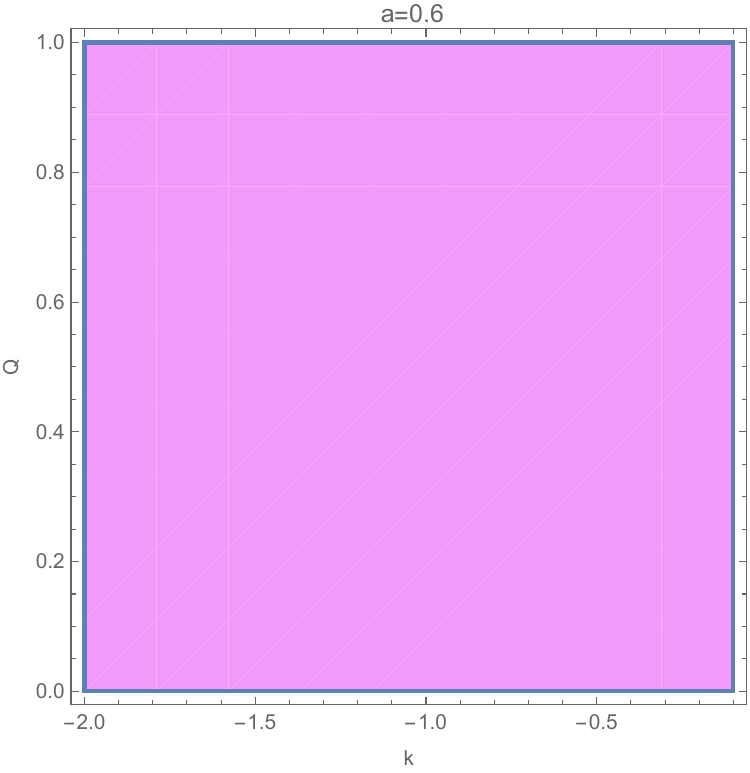}
\end{center}
\caption{Regions in the $(Q, k)$--plane, 
where the metric admits at least one real event horizon radius for  $M=1$, for the rotating solution.}
\label{3}
\end{figure}

Having discussed  the  non-rotating  solution behaviors, we now turn to the construction  of rotating  black holes within  the SRNED framework. To obtain these solutions,  the Newman-Janis algorithm is applied without a  complexification \cite{33}, which  allows   the metric to be defined as in \cite{34,35}.  This approach can be extended to  various  modified gravity theories  by incorporating additional parameters.  Within this framework,   we  investigate the physical properties of  the rotating  black hole  in SRNED backgrounds.  Adopting the Boyer-Lindquist coordinate system,  the above black hole metric can be extended to
\begin{equation}
ds^2=\left( \frac{\sigma(r)}{\Sigma(r)} -1\right) dt^2-\frac{2a^2 \sigma(r)}{\Sigma(r)}\sin^{2} \theta dtd\phi+ \left( r^2+a^2+\frac{a^2 \sigma(r) \sin^{2} \theta }{\Sigma(r)}\right)\sin^{2} \theta d\phi^2 +\frac{\Sigma(r)}{\Delta(r)}dr^2+\Sigma(r)d\theta^2,
\label{mr}
\end{equation}
where one has used
\begin{eqnarray}
\Sigma(r)=r^2+a^2 \cos^2\theta,\quad
  \Delta(r)= f(r)r^2+a^2,\quad 
 \sigma(r) = r^2- r^2f(r).  
\end{eqnarray}
In this solution,   $a$  represents  the rotation parameter of   SRNED black holes. The rotating metric is constructed under conditions ensuring it generates a physically acceptable solution of the field equations, where the associated  energy-momentum tensor is interpreted as that of an imperfect fluid rotating about the $z$-axis. To illustrate the effect of the rotation parameter on the metric function, 
 figure~(\ref{3}) depicts the regions in the $(Q, k)$--plane, for a fixed value of  $a$, in which the black hole solution admits at least one real event horizon radius. As shown in the figure, for certain values of the rotation parameter $a$, the allowed region exhibits an approximately square shape and expands the admissible parameter space compared with the non-rotating case. For other values of $a$,  the rotation has a negligible effect. In the following section, we investigate the optical  properties of such black 
holes for parameter values lying within the allowed regions.

\section{Shadow properties of SRNED black holes  }
\begin{figure}[t] 
		\begin{center}
		\centering
			\begin{tabbing}
			\centering
			\hspace{7.5cm}\=\kill
			\includegraphics[scale=.4]{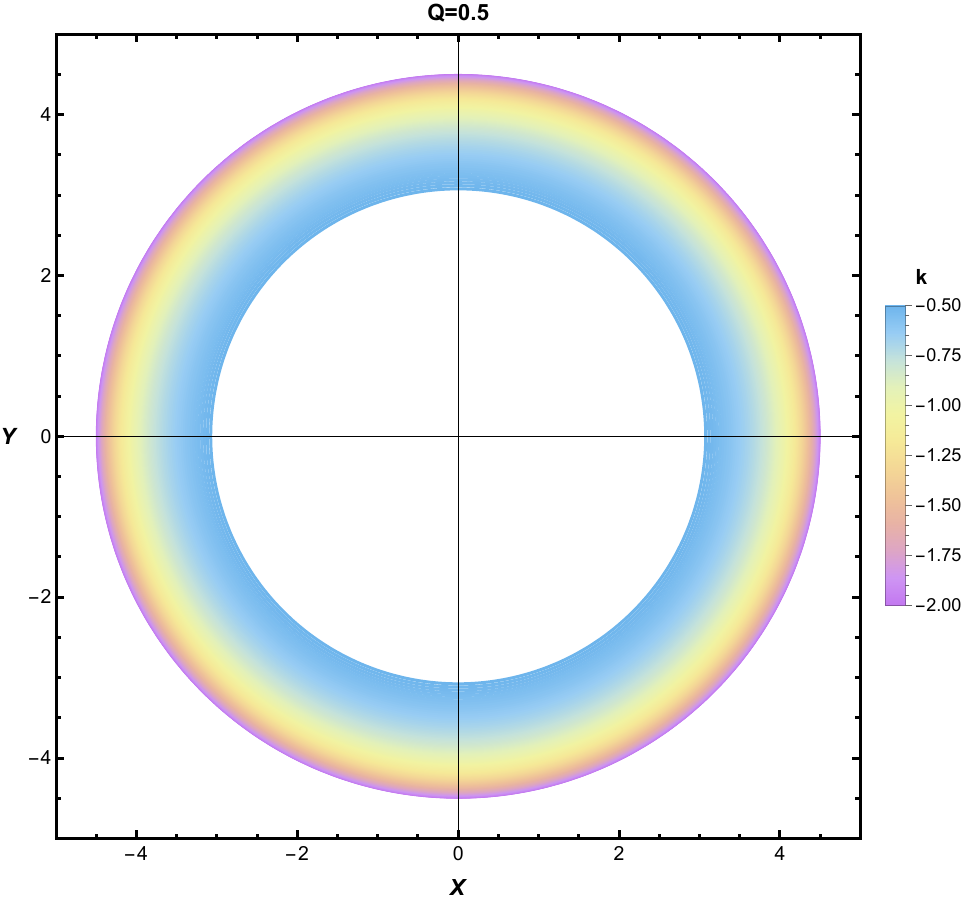} \>
			\includegraphics[scale=.4]{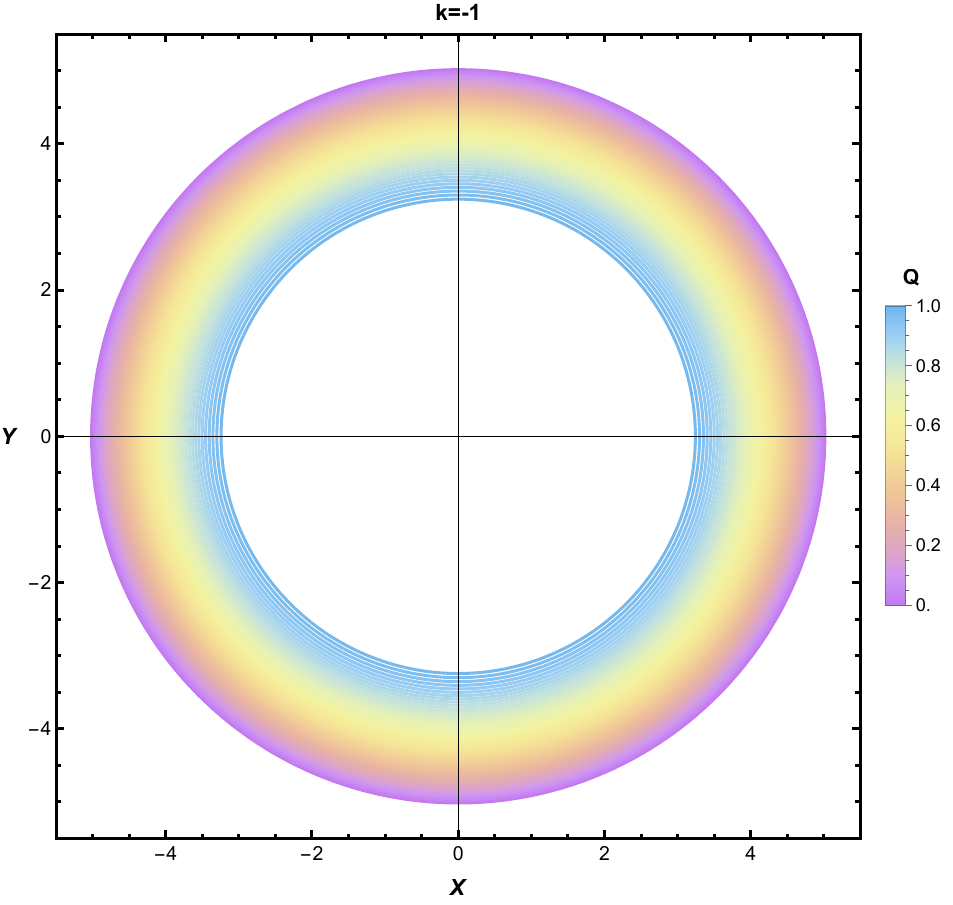} \\
		   \end{tabbing}
\caption{{\it \footnotesize  Non-rotating shadow behaviors in  the celestial coordinates $(x,y)$     by varying $Q$ and $k$ and taking  $M=1$.}}
\label{SF1}
		   \end{center}
\end{figure}

This section is devoted to investigate  the optical properties of  non-rotating and  rotating  SRNED black holes
through an elaboration  of their shadow structures. For this purpose, 
 the Hamilton--Jacobi formalism is applied using   the Jacobi action approach
\begin{equation}
\frac{\partial \mathcal{S}}{\partial \sigma} 
= -\frac{1}{2} g^{\mu \nu} 
\frac{\partial \mathcal{S}}{\partial x^\mu} 
\frac{\partial \mathcal{S}}{\partial x^\nu}.
\end{equation}
Following  \cite{36}, this action  can be decomposed as
\begin{equation}
\mathcal{S} = -Et + L\phi + S_r(r) + S_\theta(\theta)
\end{equation}
where $E$ denotes the conserved energy, $L$ represents the conserved angular momentum, 
$S_r(r)$ is a function of the radial coordinate $r$, and $S_\theta(\theta)$  is 
a function of the polar angle $\theta$.
The corresponding null geodesic equations for the radial coordinate $r$  and the angular one    $\theta$
can be derived through   a  variable separation scenario,  following  a procedure analogous to the Carter 
mechanism \cite{37}. For non-rotating SRNED black holes,  these equations take the following form
\begin{equation}
\begin{aligned}
r^2(r)\left(\frac{dS_r(r)}{dr}\right)^2 + L^2 - \frac{r^2}{f(r)}E^2 &= -\mathcal{C}, \\
\left(\frac{dS_\theta(\theta)}{d\theta}\right)^2 + L^2 \cot^2\theta &= \mathcal{C},
\end{aligned}
\label{se}
\end{equation}
where $\mathcal{C}$ is a separation constant. This leads to the following  equations of motion 
\begin{align}
\dot{t} &= -\frac{E}{f(r)}, \nonumber \\
\dot{r} &= \pm \sqrt{\mathcal{R}(r)},  \nonumber\\
r^2\dot{\theta} &= \pm \sqrt{\mathcal{C}-L^2 \cot^2\theta}, \\
\dot{\phi} &= \frac{L}{h(r)\sin^2\theta} \nonumber
\end{align}
where  $\mathcal{R}(r)$ represents a radial function given by
\begin{equation}
\mathcal{R}(r) = E^2 \left( 1 - \frac{f(r)}{r^2}(\Xi ^2+\eta)\right).
\end{equation}
  $\eta$ and $\Xi$  denote  the dimensionless impact parameters defined as follows
\begin{equation}
{{\eta}} = \frac{\mathcal{C}}{E^2}, \qquad  { \Xi} = \frac{L}{E}.
\end{equation}
In this way, the shadow boundary can be determined by considering the conditions for unstable circular photon 
orbits, namely
\begin{equation}
\mathcal{R}(r)\big|_{r=r_o} = 0, \qquad 
\frac{\mathcal{R}(r)}{dr}\big|_{r=r_o} = 0,
\label{sheq}
\end{equation}
where $r_o$ represents the radius of the circular photon orbit.  The above impact parameters can be determined by solving  Eq. (\ref{sheq}). The calculation gives
\begin{equation}
\Xi ^2+\eta=\frac{3 r^{4} k}{2 Q \sqrt{1-\frac{k Q}{r}}\, \left(k  Q -r \right)+3 k  r \left(r -2 M \right)}.
\end{equation}
Taking $ Q = 0$, we recover the circular algebraic equation
\begin{equation}
\Xi ^2+\eta=27 \,M^2
\end{equation}
 associated with the  ordinary Schwarzschild black hole \cite{MM}. The observed shape of the black hole shadows, as noticed at spatial infinity, 
can be characterized by the celestial coordinates $(X,Y)$ being 
\begin{equation}
\begin{aligned}
X &= \lim_{r_{ob}\to\infty}\left(-r_{ob}^2 \sin\theta_{ob}\,\frac{d\phi}{dr}\right), \\
Y  &= \lim_{r_{ob}\to\infty}\left(r_{ob}^2 \,\frac{d\theta}{dr}\right),
\end{aligned}
\end{equation}
where $r_{ob}$ represents the distance of the observer from the black hole and 
$\theta_{ob}$ is the inclination angle between the observer line of sight and 
the symmetry axis of the black hole. The coordinate $X$ corresponds to the 
apparent perpendicular transfer from the axis of symmetry, while $Y$ 
assesses the movement from the projection onto the equatorial plane. For 
null geodesics, the shadow boundary satisfies the following algebraic equation 
\begin{equation}
X^2 + Y^2 = \Xi^2 + \eta.
\end{equation}
In figure (\ref{SF1}), we show the shadow behaviors by examining the effects of the charge $Q$ and the SRNED parameter $k$, by fixing the mass $M=1$. 

\begin{figure}[!ht]
		\begin{center}
		\centering
			\begin{tabbing}
			\centering
			\hspace{2.cm}\=\kill
			\includegraphics[scale=0.52]{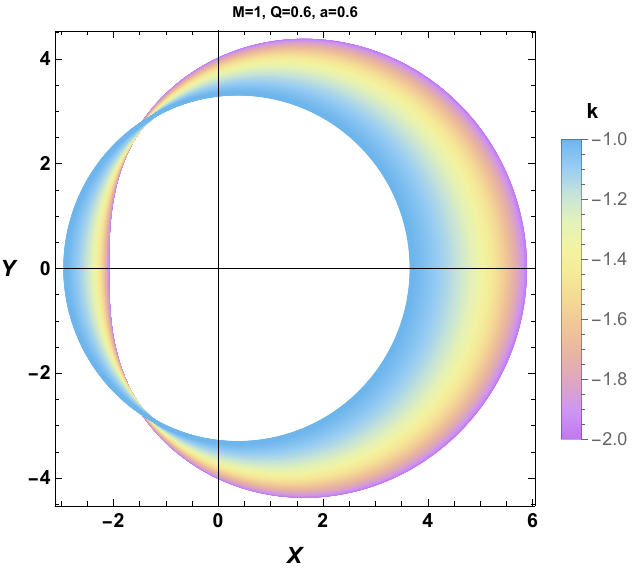} 
	\hspace{0.1cm}		\includegraphics[scale=0.5]{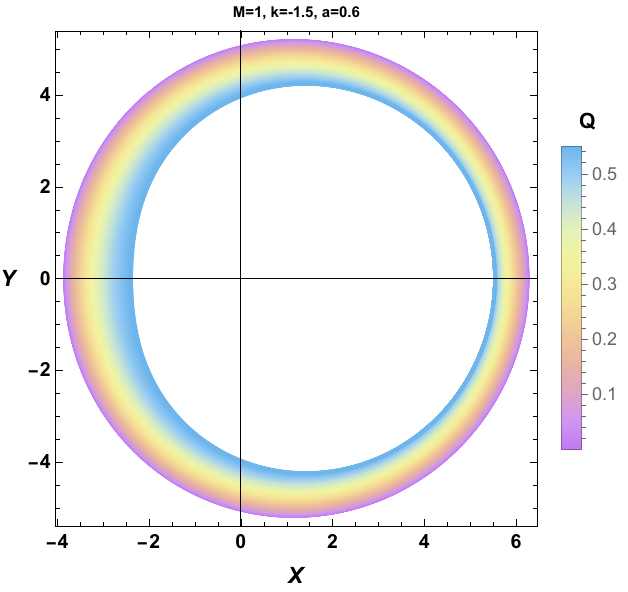}\hspace{0.1cm}	\includegraphics[scale=0.54]{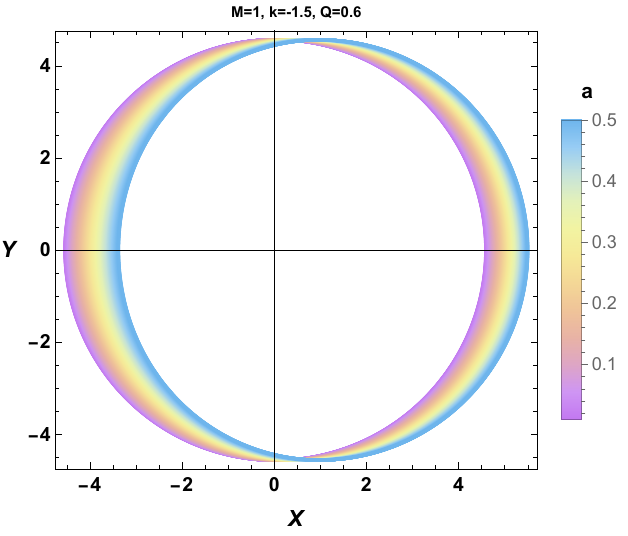}\\ 
	
		   \end{tabbing}
\caption{ \it \footnotesize Black hole shadow behaviors  in terms of the charge $Q$, the rotation $a$, and the SRNED parameter $k$ are varied  with $M=1$}
\label{rsa}
\end{center}

\end{figure}

It is evident  from the figure that both the charge and the SRNED parameter have a similar effect on the black hole shadow. In this way, increasing either $Q$ or $k$ reduces the shadow size without affecting its shape. This proves that the combined effect of the charge and  the SRNED  parameter controls the size of the  curve shadows. After that,  we  inspect the effect of the rotation parameter on the reconsidered black   solutions by examining how the moduli  space controls both the size and  the shape of the resulting shadows.
To  do so, we start from the metric (\ref{mr}) and apply the Hamilton–Jacobi separation method. For this black hole solution, the resulting   equations of motion can be expressed as follows
\begin{align}
\Sigma \dot{t}& = \frac{r^2+a^2}{\Delta}\left[ E\left( r^2+a^2\right) -a L\right] 
+a\left[  L-aE\sin^2\theta \right]  \\
(\Sigma \dot{r})^2 &=\mathcal{R}(r) \\
( \Sigma\dot{ \theta})^2 & =\Theta(\theta)\\
\Sigma \dot{\phi} &= \left[ L \csc ^2 \theta-aE\right] 
+\frac{a}{\Delta} \left[ E\left(r^2+a^2\right) -aL\right],
\end{align}
where $ \mathcal{R}(r)$  and $\Theta(\theta) $  are radial functions given by
 \begin{align}
\mathcal{R}(r) &= \left[ E \left( r^2 + a^2 \right) - a L \right]^2 
- \Delta \left[ \mathcal{C} + \left( L - a E \right)^2 \right], \\
\Theta(\theta) &= \mathcal{C} - \left( L \csc \theta - a E \sin \theta \right)^2 
+ \left( L - a E \right)^2.
\end{align}
Assuming the unstable spherical photon orbits and projecting the null trajectories onto the observer celestial plane, the apparent shadow boundary is controlled by the two impact parameters 
 \begin{align}
\eta 
&= \frac{r^{2}\!\left[\,16 a^{2}\Delta(r) + 8 r\,\Delta(r)\Delta'(r) - 16 \Delta(r)^{2} - r^{2}\Delta'(r)^{2}\right]}{a^{2}\,\Delta'(r)^{2}}\bigg|_{r=r_0}, \\[4pt]
\Xi 
&= \frac{(r^{2}+a^{2})\,\Delta'(r) - 4 r\,\Delta(r)}{a\,\Delta'(r)}\bigg|_{r=r_0}.
\end{align}
In figure (\ref{rsa}), we show  the black hole shadow  behaviors  in terms of the moduli space parameter. 

\begin{figure}[!ht]
    \centering
    \includegraphics[scale=0.3]{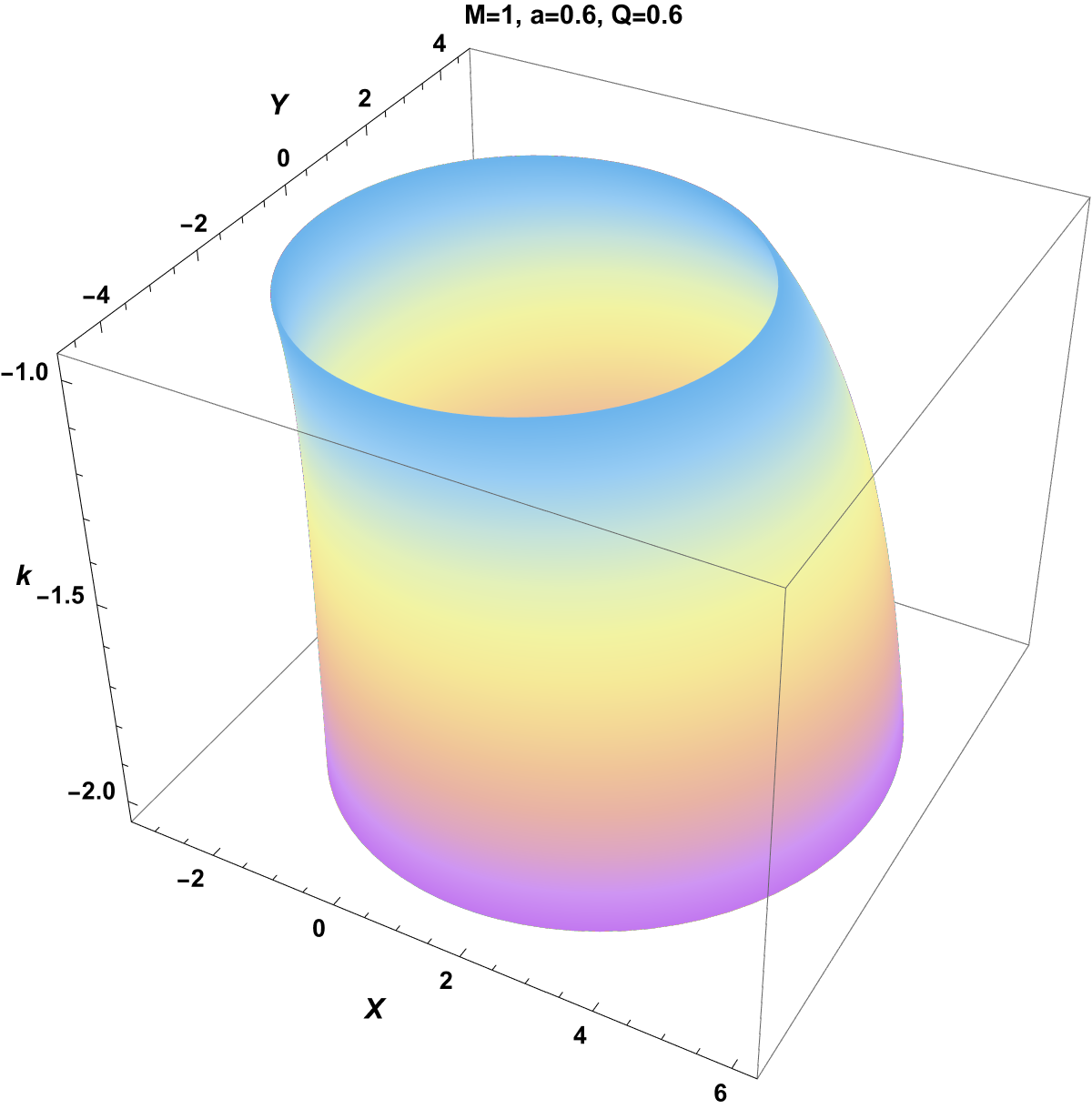}
    \includegraphics[scale=0.3]{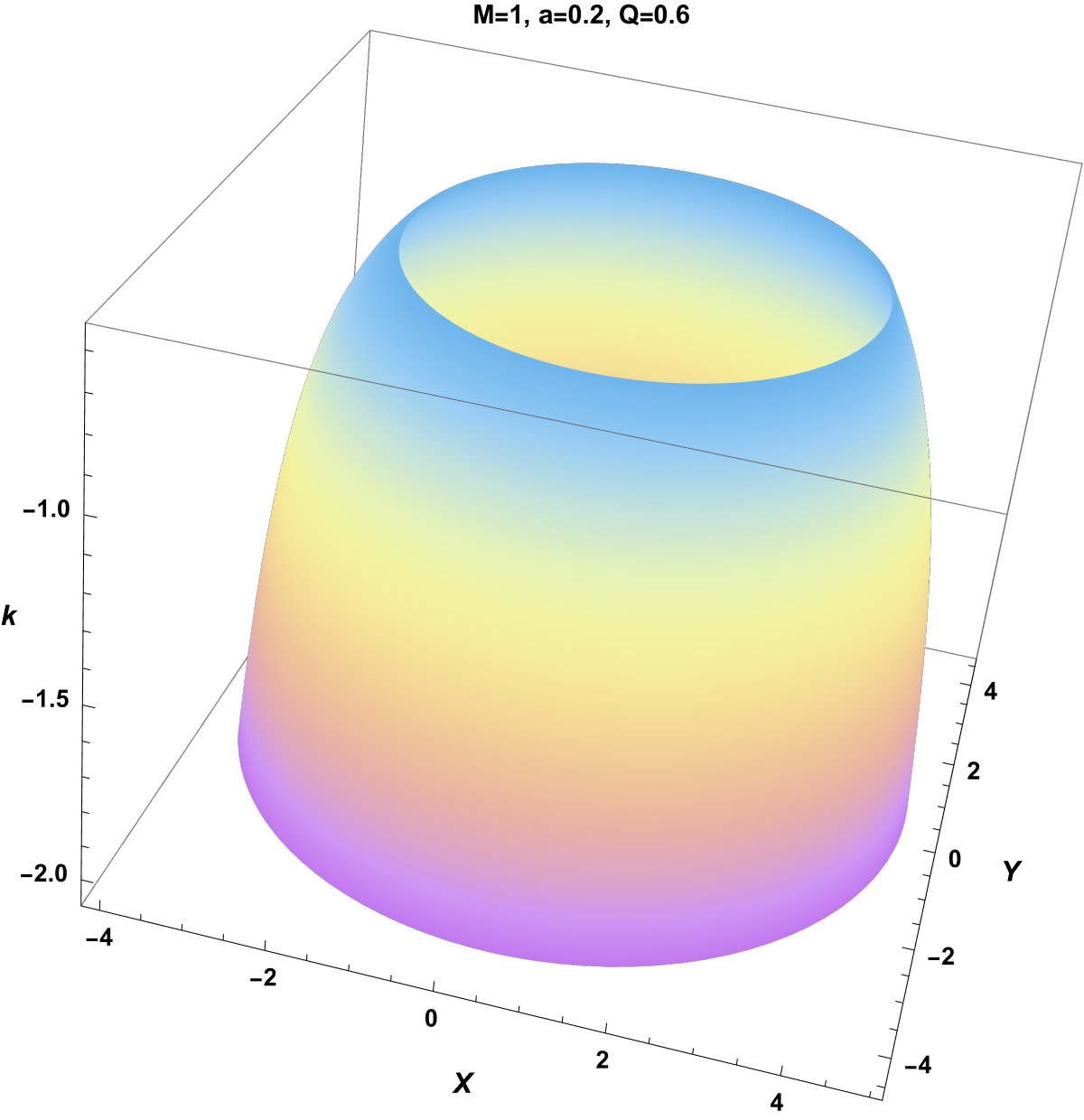}
    \caption{\footnotesize Behavior of the black hole shadow for different values of the SRNED parameter $k$, with the rotation parameter fixed at $a=0.6$ (left) and $a=0.2$ (right) in three. dimensions.}
    \label{rsa1}
\end{figure}

Precisely, we focus on the effects of the charge $Q$, the rotation parameter $a$, and the SRNED parameter $k$. The rotation parameter $a$   generates a  similar effect to that in ordinary rotating black holes, rather reducing the shadow size and deforms the   shape into a D-like structure. This indicates its role as a deformed parameter in these black hole solutions. Considering the charge, increasing its value reduces the shadow size and affects its shape. Nevertheless, the SRNED parameter $k$ displays  an interesting effect on the shadow shape. Indeed as $k$ decreases, the shadow slowly develops a D-like deformation.  These interesting features motivate one to further investigate  three dimensions  shadow profiles. By considering $k$ parameter values as an extra direction, this behavior is shown  in figure~(\ref{rsa1}).  In this figure, we examine the influence of the SRNED parameter $k$ on the  shadow configurations in  three dimensional representation  for two different values of the rotation parameter, $a=0.2$ and $a=0.6$.
 
It is evident from the figure that both surfaces exhibit a similar overall structure. However, increasing  the rotation parameter modifies their geometry, leading to  a  greater deformation for $a=0.6$.  For $a=0.2$,  conversely, the surface exhibits a more uniform geometry  closely resembling a cylindrical shape.

\section{EHT observational constraints on black hole parameters}

In order to establish a bridge between the   theoretical finding and  the empirical data, we present in the following section  an examination of  rotating black holes in the SRNED model by considering the  EHT  observational data \cite{EH1,EH2,EH3}. Especially, we employ the EHT data for M$87^*$ and Sgr~A$^*$ to constrain on the corresponding  black hole  parameters according the investigations reported in ~\cite{EventHorizon,Chakhchi,Gogoi,m1}. The corresponding constraints are obtained by evaluating  the fractional deviation of the shadow diameter relative to  a Schwarzschild black hole defined as follows
\begin{equation}
{ d} = \frac{R_s}{r_{sh}}-1,
\end{equation}
where $R_s$ represents the shadow radius. $r_{sh}$ corresponds to the shadow radius of the Schwarzschild black hole. The dimensionless ratio $R_s/M$ provides  an important  observable to compare theoretical predictions with observational measurements. The 1-\(\sigma\) and 2-\(\sigma\) confidence intervals derived from  EHT observations are reported in Table~\ref{t1}.

\begin{figure}[htbp]
   \includegraphics[scale=0.3]{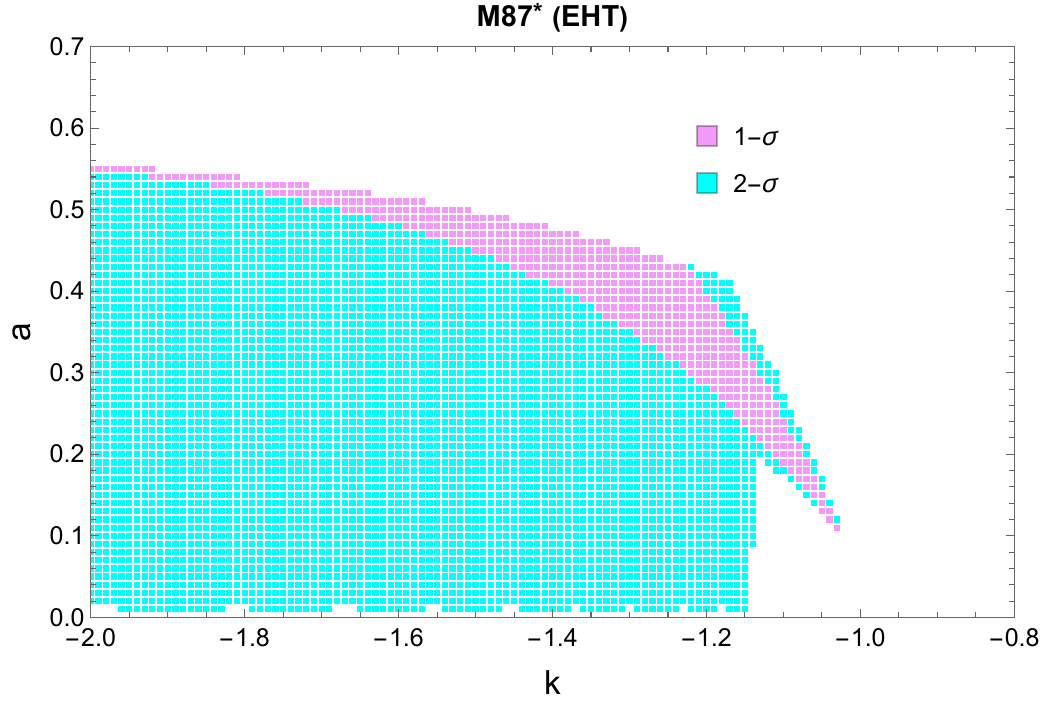}%
   \hspace{2mm}%
    \includegraphics[scale=0.3]{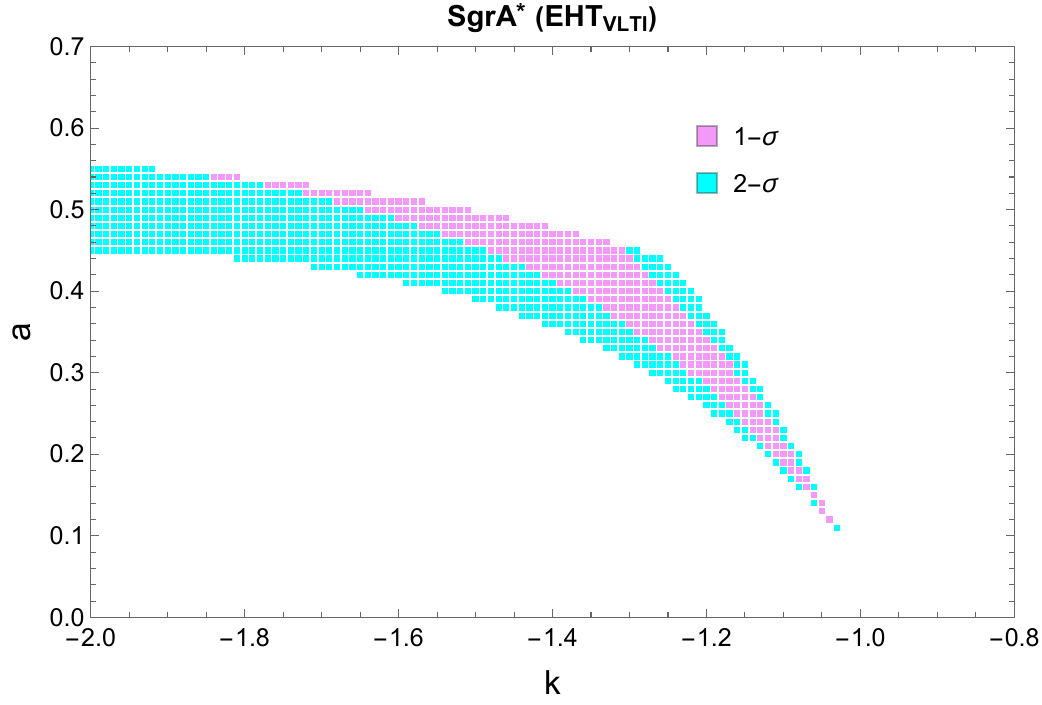}    
   \hspace{2mm}%
    \includegraphics[scale=0.3]{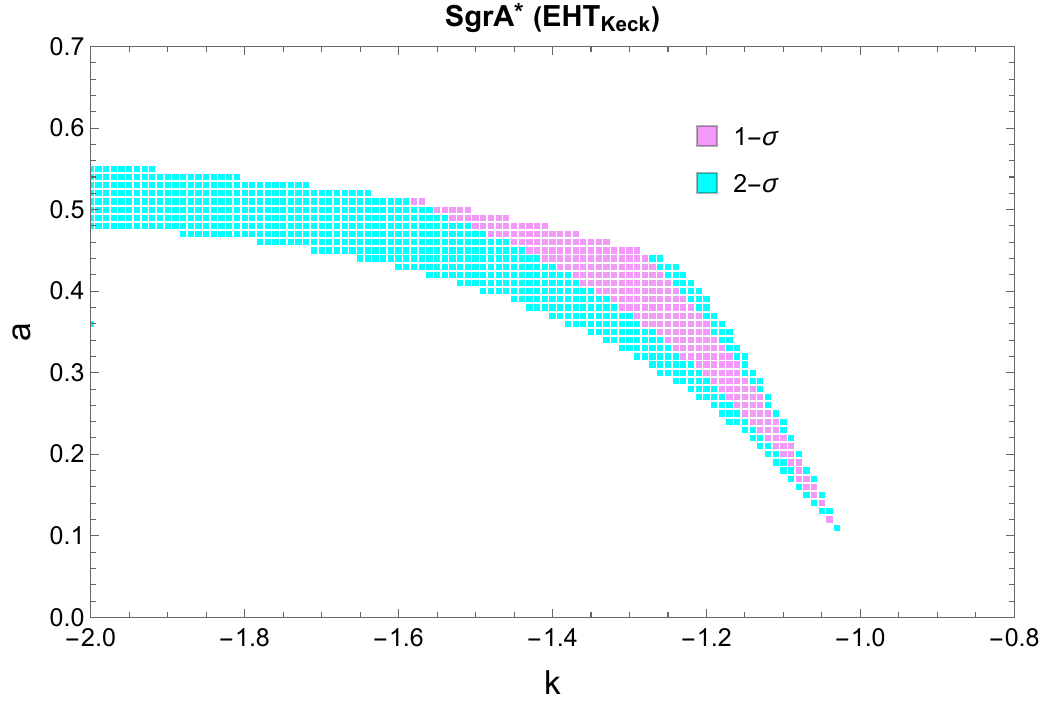}%
                           
\caption{ \it \footnotesize Constraint regions in the $( a,k)$ plane obtained from CUDA-based simulations, showing agreement with the EHT observations of M$87^*$ and Sgr~A$^*$ within $1-\sigma$ and $2-\sigma$ confidence levels for $Q=0.4$  with $M=1$.}

  \label{fig:Moduli}
\end{figure}

\begin{table}[h!]
\centering
\begin{tabular}{|c|c|c|c|}
\hline
{Black Hole} & {Deviation (\(d\))} & {1-\(\sigma\) Bounds} & {2-\(\sigma\) Bounds} \\ 
\hline
M87$^*$ (EHT) & $-0.01^{+0.17}_{-0.17}$ & $4.26 \leq \frac{R_s}{M} \leq 6.03$ & $3.38 \leq \frac{R_s}{M} \leq 6.91$ \\ 
\hline
Sgr~A$^*$ (EHT$_{\text{VLTI}}$) & $-0.08^{+0.09}_{-0.09}$ & $4.31 \leq \frac{R_s}{M} \leq 5.25$ & $3.85 \leq \frac{R_s}{M} \leq 5.72$ \\ 
\hline
Sgr~A$^*$ (EHT$_{\text{Keck}}$) & $-0.04^{+0.09}_{-0.10}$ & $4.47 \leq \frac{R_s}{M} \leq 5.46$ & $3.95 \leq \frac{R_s}{M} \leq 5.92$ \\ 
\hline
\end{tabular}
\caption{ \it \footnotesize Estimated fractional deviations and corresponding bounds for M87$^*$ and Sgr~A$^*$ black holes.}
\label{t1}
\end{table}
In  this work, we perform numerical calculations  to identify the pairs $(a, k)$ that produce shadow configurations compatible with the observational results within the $1-\sigma$ and $2-\sigma$ confidence intervals reported by the EHT collaboration.  In particular, fixing the charge  $Q = 0.4$, the corresponding results are illustrated in figure (\ref{fig:Moduli}).
 
As shown in the three panels of this  figure, the density of points aligned with the empirical data increases for small values of $a$ and $k$.  This result demonstrates that the moduli space of the black hole can successfully reproduce observational signatures. To demonstrate the relationship between the two parameters, we fix one parameter while constraining the other. Taking $k = -1.3$ and $Q=0.4$, Table~(\ref{taaa}) offers the corresponding ranges of $a$ that provides coherence with the $1-\sigma$ and $2-\sigma$  intervals for the three experimental scenarios.

\begin{table}[h!]
\centering
\begin{tabular}{|c|c|c|c|}
\hline
Confidence interval & M87$^*$ (EHT) & Sgr~A$^*$ (EHT$_{\text{VLTI}}$) & Sgr~A$^*$ (EHT$_{\text{Keck}}$) \\ 
\hline

$1-\sigma$ & $ 0.36\leq a\leq 0.45 $ & $ 0.36\leq a\leq 0.44 $& $ 0.38\leq a\leq 0.45 $ \\ 

\hline

$2-\sigma$ & $ 0.01\leq a\leq 0.45 $ & $ 0.32\leq a\leq 0.44 $& $ 0.33\leq a\leq 0.45 $ \\

\hline
\end{tabular}
\caption{ \it \footnotesize Constraints on the rotating parameter $a$ at $k = -1.3$ from EHT empirical  data.}
\label{taaa}
\end{table}
The table reveals that only particular values of $a$ generate results in agreement with the empirical evidence. Particularly, $a$ should be smaller than $0.55$, when all other parameters are fixed. 
\section{SRNED black holes in machine learning scenarios}
In this section, we  reconsider  the study of the optical behaviors by   investigating whether the regions of the moduli space  
 consistent with the EHT observations using machine learning methods.  The latter offers a suitable alternative by learning the  relation between the model parameters and the empirical constraints on the generated black hole   shadows \cite{i6,i7,i8,i9}. In this regard, a fully connected neural network (FCNN) is used to classify the $(a,k)$ moduli space based on  either the rotating black hole results are compatible with the EHT observations.

Implementing a numerical simulation, we have  generated the datasets  by analyzing the $(a,k)$ moduli space and comparing the corresponding shadow with the empirical constraints of $M87^*$, Sgr A$^*_{\mathrm{VLTI}}$, and Sgr A$^*_{\mathrm{Keck}}$ within the  $1-\sigma$ and $2-\sigma$ bounds. The resulting parameter configurations are grouped into two classes compatible with the observational constraints. Configurations satisfying the corresponding  EHT shadow bounds are classified  as Class 1, while Class 0 denotes configurations incompatible with the EHT observations. Unphysical and redundant configurations are discarded. The input parameters are then standardized, and the resulting dataset is divided into training, validation, and test sets using a fixed random seed, \texttt{random\_state = 42},  to ensure the reliability of the analysis.

\begin{figure}[!ht]
 \centering
  \includegraphics[scale=0.25]{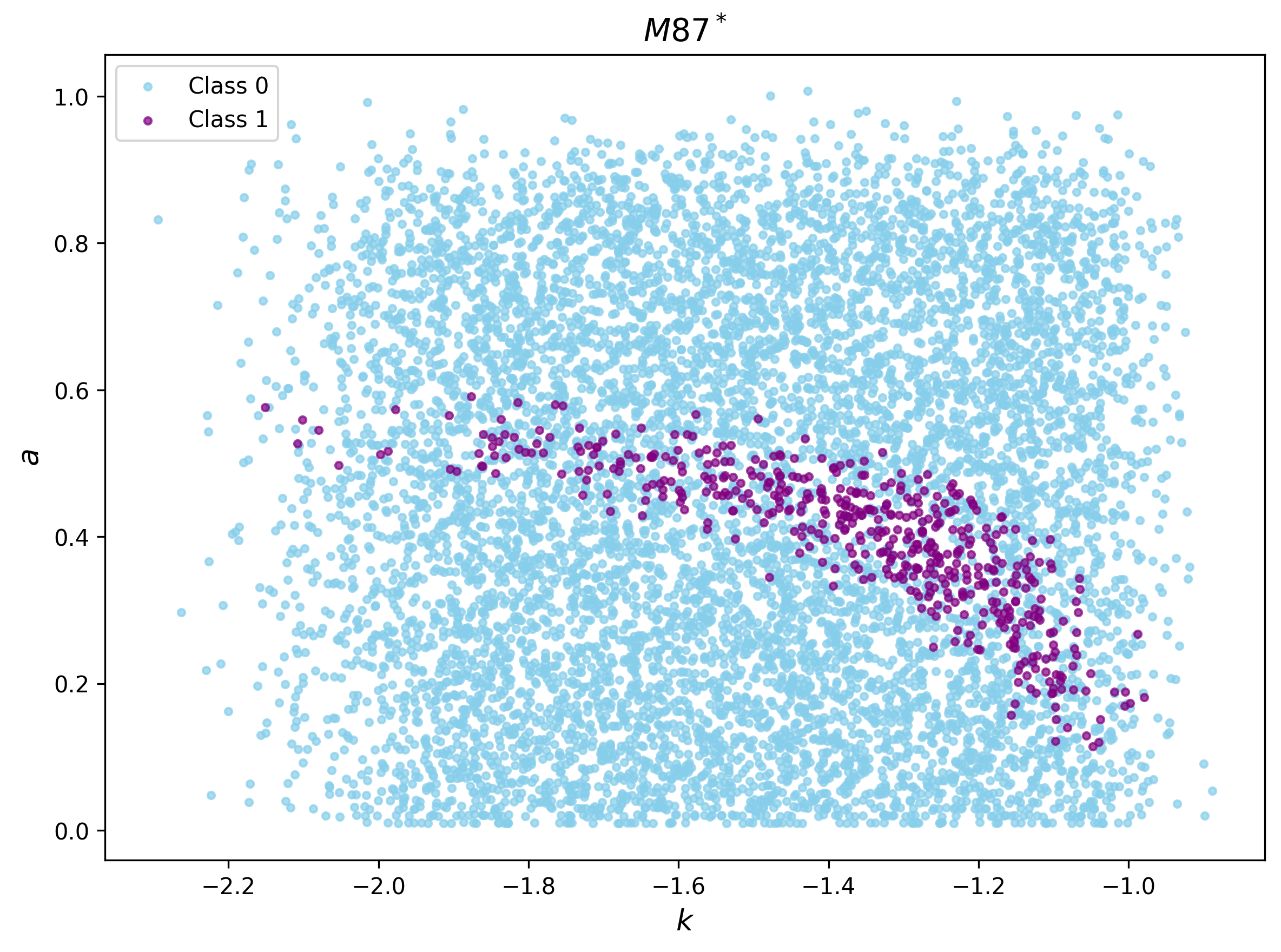}\hspace{2mm} 
  \includegraphics[scale=0.25]{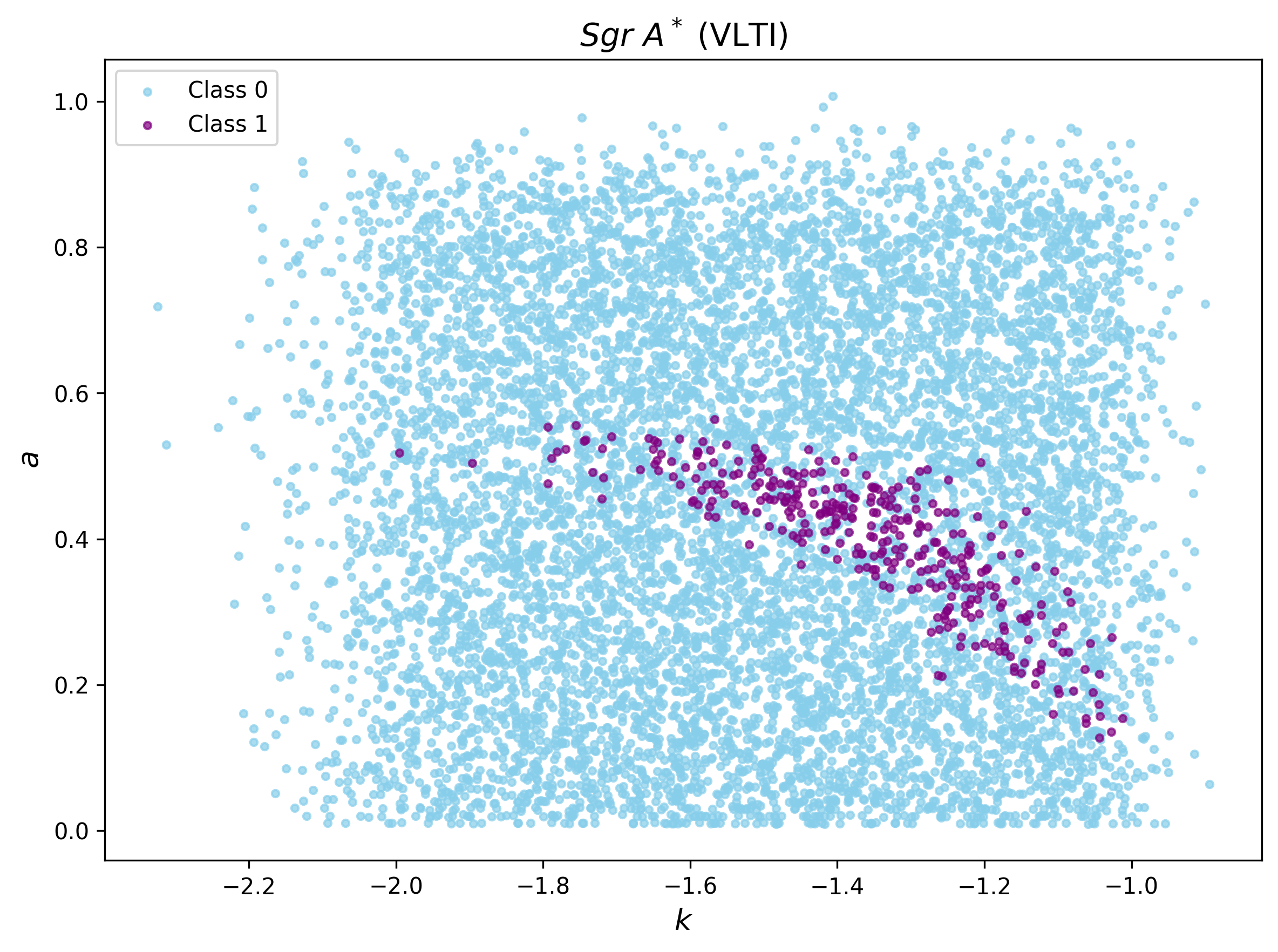}\hspace{2mm} 
  \includegraphics[scale=0.25]{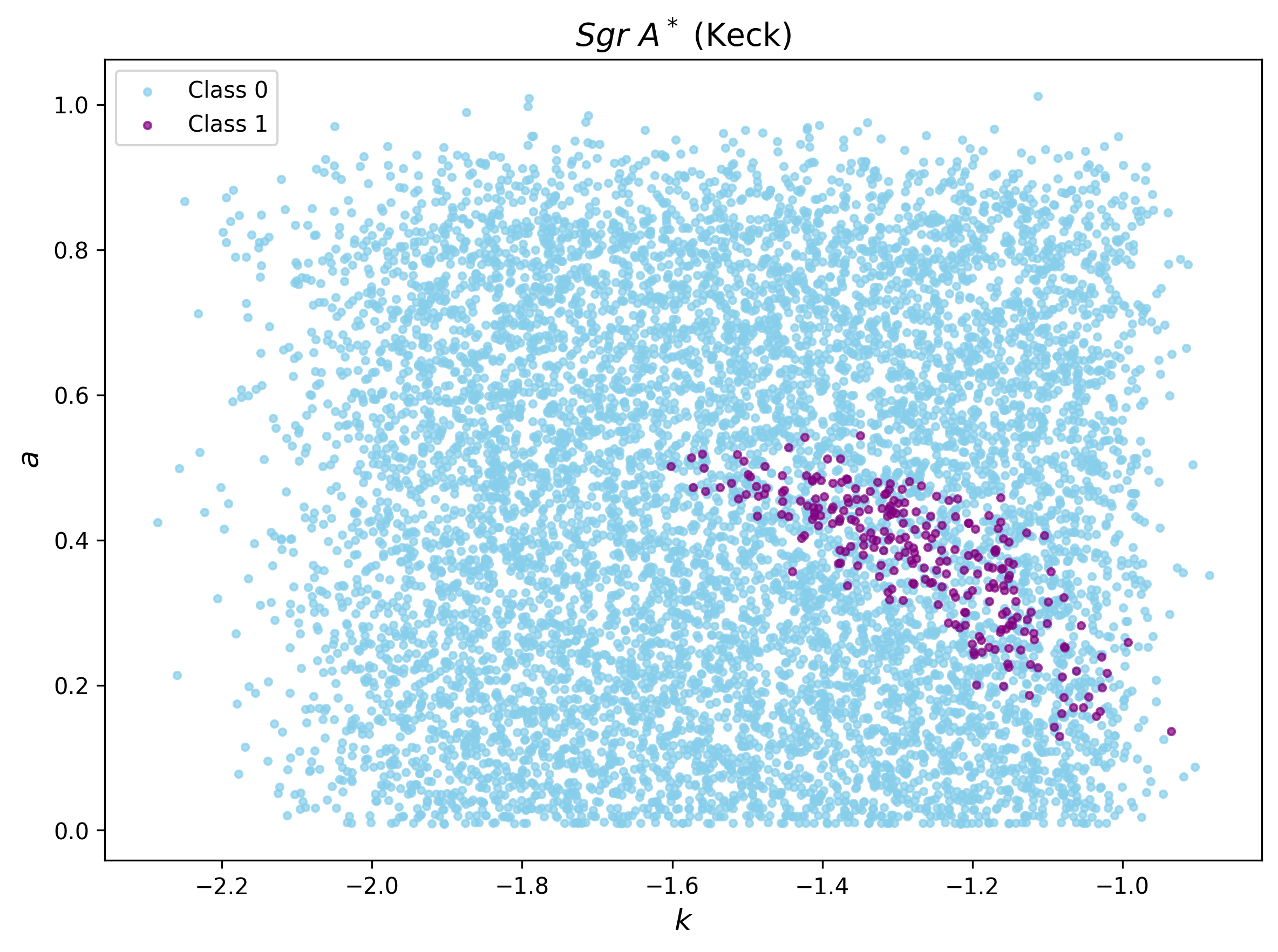} \vspace{2mm} 
  \includegraphics[scale=0.25]{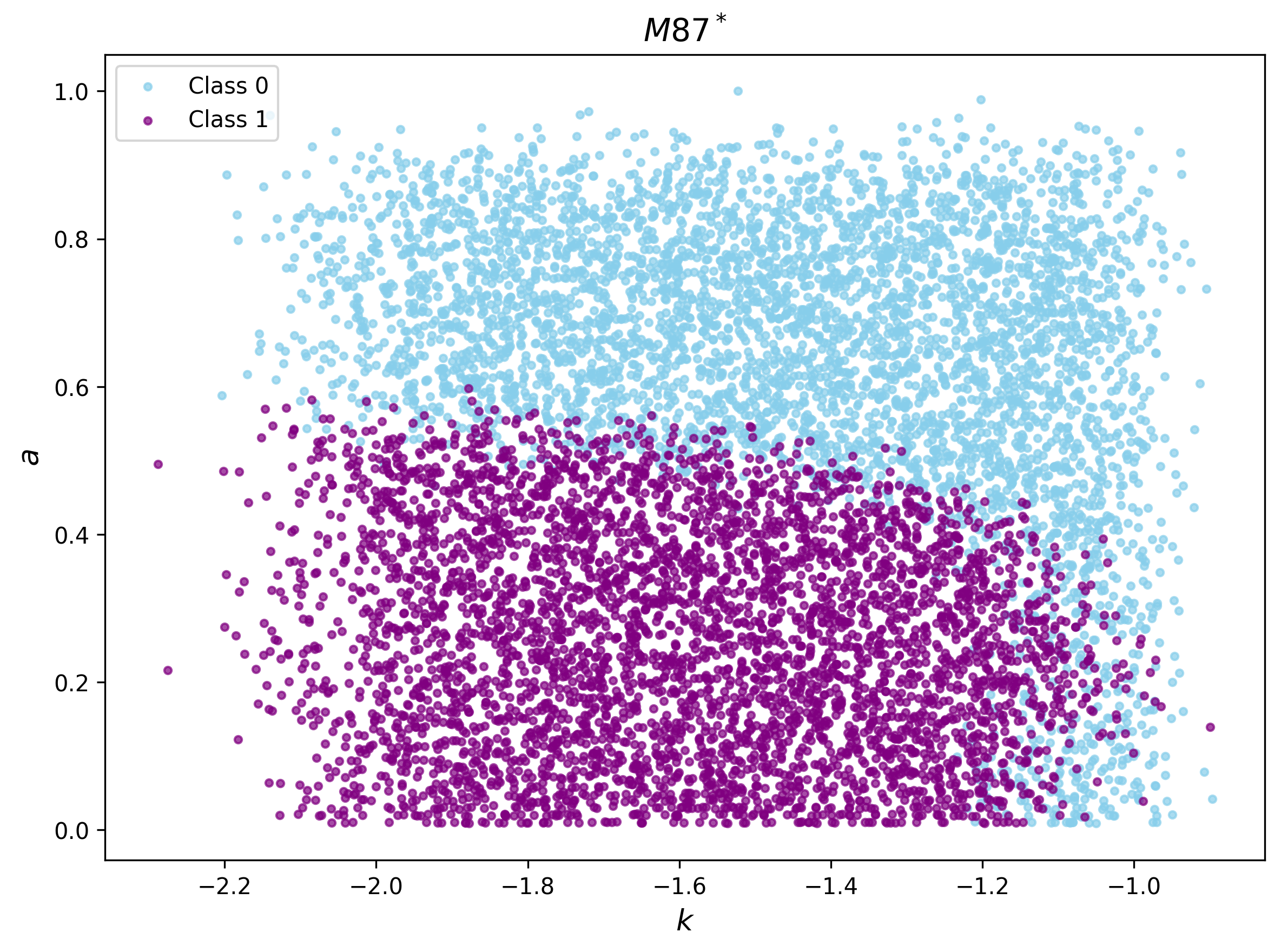}\hspace{2mm} 
  \includegraphics[scale=0.25]{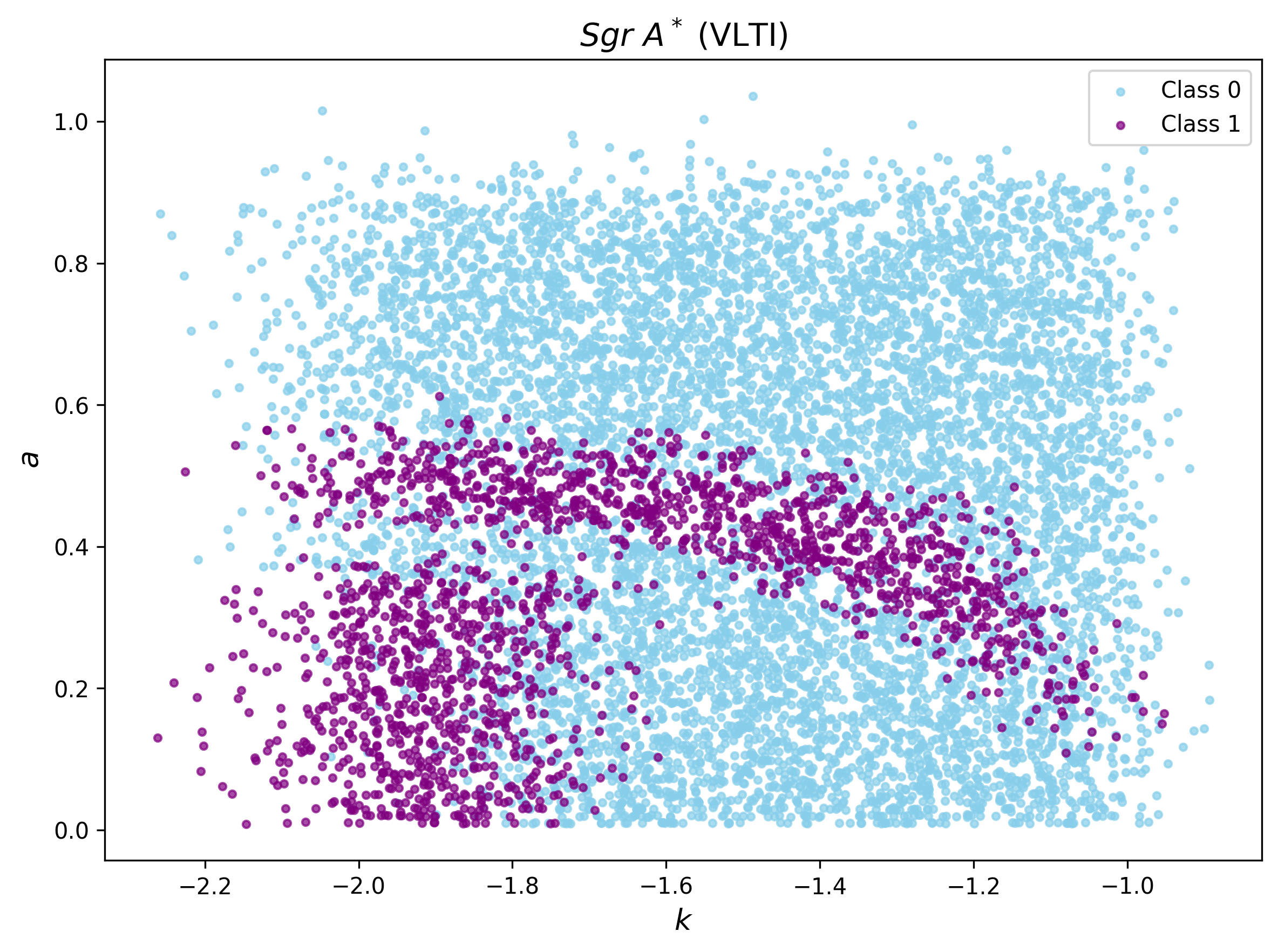}\hspace{2mm} 
  \includegraphics[scale=0.25]{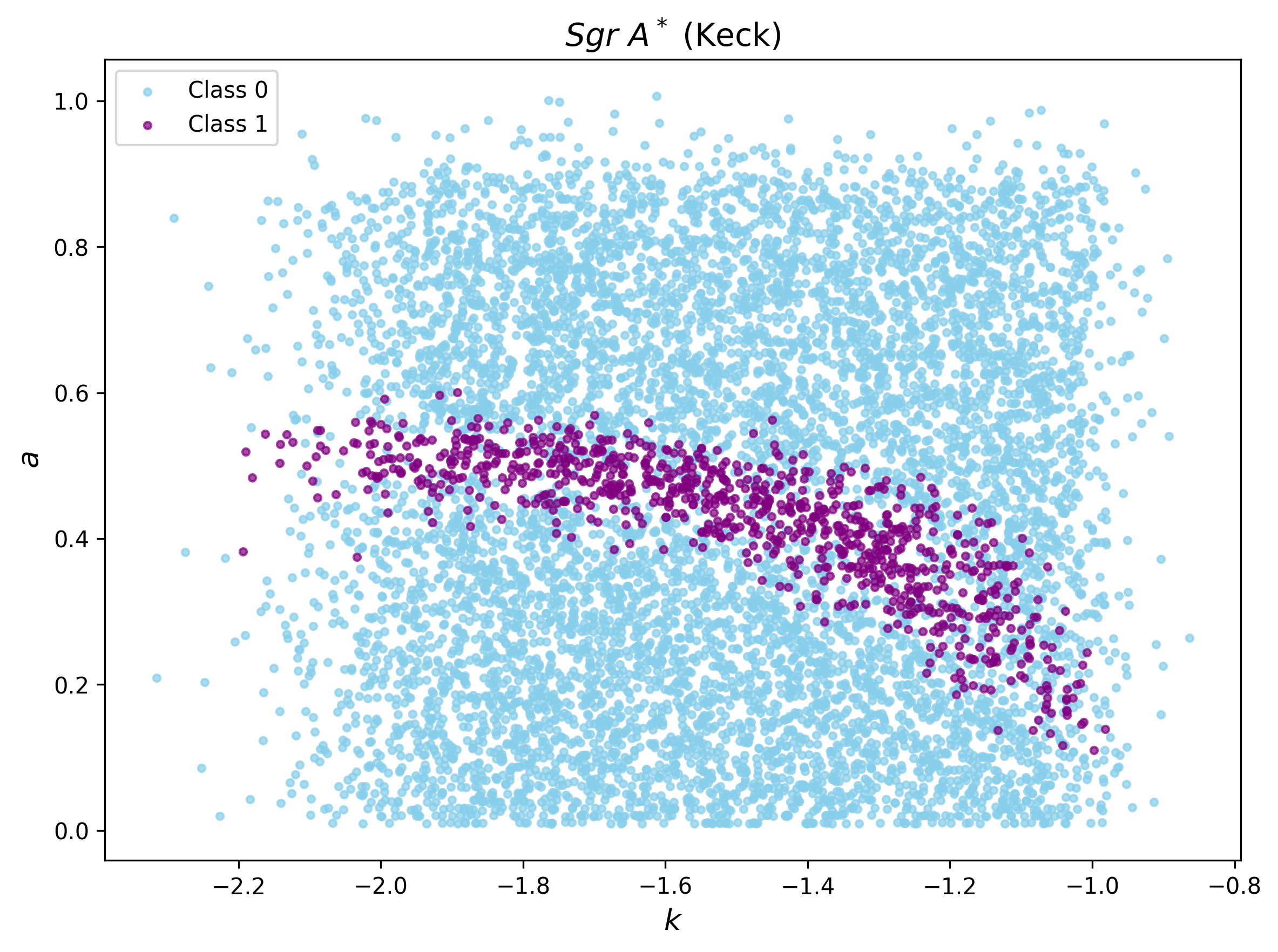} \caption{Distributions of the generated datasets for $M87^*$, $Sgr A^*_{\mathrm{VLTI}}$, and $Sgr A^*_{\mathrm{Keck}}$ at the $1-\sigma$ (top row) and $2-\sigma$ (bottom row) confidence levels.} \label{fig}
 \end{figure}

The resulting parameter distributions are illustrated in figure~(\ref{fig}). For each observational dataset, the compatible and incompatible configurations are distributed across separated regions of the $(a,k)$ parameter space, separated by nonlinear boundaries. The position and the shape of these regions depend on both the observational dataset and the adopted confidence level. This  distribution reveals the presence of nonlinear correlations between the model  parameters, making the problem well suited to a supervised machine learning approach.

The trained FCNN has been evaluated using an independent test set, with its accuracy assessed separately for each observational dataset. The results demonstrate that the network effectively distinguishes between   parameter configurations satisfying the EHT constraints to those do not.  This demonstrates that the FCNN can determine the corresponding selection boundary in the reduced moduli space $(a,k)$. The implementation of the black hole models for M87$^*$, Sgr$A^*_{\mathrm{VLTI}}$, and Sgr$A^*_{\mathrm{Keck}}$ at the $1-\sigma$ and $2-\sigma$ bounds is shown in Table~\ref{tab1}.

\begin{table}[!ht]
\centering
\caption{Model performance of the FCNN model for the $1-\sigma$ and $2-\sigma$ observational constraints.}
\label{tab1}
\begin{tabular}{l|c|c}
\toprule
 & $1-\sigma$ & $2-\sigma$ \\
 \midrule
{Metric} 
& M87$^*$ \quad SgrA$^*_{\mathrm{VLTI}}$ \quad SgrA$^*_{\mathrm{Keck}}$
& M87$^*$ \quad SgrA$^*_{\mathrm{VLTI}}$ \quad SgrA$^*_{\mathrm{Keck}}$ \\
\midrule
Test Accuracy 
& 0.9448 \quad 0.9651 \quad  0.9673
& 0.9459\quad 0.8958 \quad  0.9358\\
\bottomrule
\end{tabular}
\end{table}

To inspect the robustness of the supervised model, we use a Monte Carlo voting method based on Gaussian perturbations of the accepted input data. For each sample in the separate test set, $20$ perturbed sample are developed by including Gaussian noise with a standard deviation of $0.01$ to the normalized input data. Each perturbed sample is then independently treated by the FCNN, and the final prediction is obtained through majority voting among the $20$ outputs. Since the perturbations are introduced only after the training phase, they do not affect the target labels generated by the numerical simulations. This approach allows  the stability of the classifier around each test sample to be assessed while reducing its sensitivity to small numerical fluctuations.

The confusion matrices obtained for the M87$^*$, SgrA$^*_{\mathrm{VLTI}}$, and Sgr$A^*_{\mathrm{Keck}}$ black hole models under the $1-\sigma$ and $2-\sigma$ observational constraints  are reported in Table~(\ref{c11}). The results demonstrate that the FCNN correctly classifies nearly all samples, with only a very small number of misclassifications.

\begin{table}[!ht]
\centering
\caption{Confusion matrix for FCNN model  with  $1-\sigma$ and $2-\sigma$ observational constraints.}
\label{c11}
\begin{tabular}{l|l|c|c|c}
\toprule
& Class & {M87$^*$} 
& {SgrA$^*_{\mathrm{VLTI}}$} & {grA$^*_{\mathrm{Keck}}$} \\
\midrule
\multirow{2}{*}{$1-\sigma$} 
&  & Class 0 \quad Class 1 &  Class 0 \quad Class 1 &  Class 0 \quad Class 1 \\
& Class 0 &1647 \qquad  30&1696\qquad 13 &1712 \qquad 15  \\
& Class 1 &68 \qquad 32& 49 \qquad 19 &43 \qquad 7  \\
\midrule
\multirow{2}{*}{$2-\sigma$} 
&  & Class 0 \quad Class 1 &  Class 0 \quad Class 1 &  Class 0 \quad Class 1 \\
& Class 0 &843 \qquad 50&1332\qquad 80 &1551 \qquad 43 \\
& Class 1 &46 \qquad 838 &105 \qquad 260&71\qquad 112 \\
\bottomrule
\end{tabular}
\end{table}
The results indicate that the  classifier achieves higher accuracy under the $1-\sigma$ observational constraints compared with the $2-\sigma$ constraints. Among the three datasets, the 	Sgr$A^*_{Keck}$ case consistently yields the highest classification accuracy.  However, the FCNN maintains strong performance across all datasets, highlighting its ability to generalize effectively across different confidence levels of the EHT observational constraints.

\section{Conclusions}
Motivated by a recent work on the thermodynamic properties of a new class of modified black holes in SRNED frameworks, we have examined  the shadow geometrical configuration of rotating and non-rotating SRNED black holes and their dependence on the associated moduli space. More specifically, we have shown that, in the non-rotating case, the charge $Q$ and  the SRNED parameter $k$ have  a similar effect on the shadow size without affecting its shape. For rotating solutions parameter, however, we have observed that the charge, the SRNED parameter, and the rotating parameter influence the size and the shape for the shadow curves. Precisely, the  SRNED  parameter has generated an interesting effect. Considering the $k$ parameter as an extra direction,  we have shown such an effect in 3-dimensional graphical representations. Furthermore, we have identified the corresponding allowed regions in the parameter space by confronting the theoretical predictions with the empirical predictions constraints provided by EHT collaboration. Using this framework, we have shown that,  for negative values of the SRNED parameter, the rotating parameter $a$ should be smaller than $0.55$ in order to match such predictions with EHT observations.  To complete this analysis, we have employed machine learning methods to examine the compatibility of the reconsidered black hole with EHT observations and to identify the associated regions of the moduli space. The results have demonstrated a compatibility with the  Sgr~A$^*_{Keck}$  black hole model.    

This study could be extended to investigate the optical properties of these black holes in the presence of a cosmological constant, as well as to explore other observational signatures, including gravitational lensing and energy emission rates. Moreover, future investigations could focus on applying more advanced machine learning techniques to further explore the parameter space and deepen the understanding of the optical properties of  the SRNED black holes. These issues could be  addressed in future works. 
\section*{Acknowledgements}
The author  would like to thank S. E. Baddis,   A. Belhaj,  H. Belmahi,  S. E. Ennadifi and M. Nach  for   recent collaborations, discussions,  and encouragement.     She  gratefully acknowledges the financial support of the CNRST in the
frame of the PhD Associate Scholarship Program PASS.

\end{document}